# Multipartite entangled states with two bosonic modes and qubits[*]

P. P. Munhoz and F. L. Semião

*Departamento de Física, Universidade Estadual de Ponta Grossa - Campus Uvaranas, 84030-900 Ponta Grossa, PR, Brasil*

We theoretically investigate the role of different phases of coupling constants in the dynamics of atoms and two cavity modes, observing deterministic generation of prototype or hybrid Bell, W, GHZ, and cluster states. Commonly induced dipole-dipole interactions (far-off resonance) are inhibited between particular pairs of qubits under suitable choice of those phases. We evaluate the generation fidelities when imperfections such as dissipative environments and time precision errors are considered. We show violation of local realism for the generated cluster state under such imperfections, even when approaching the weak coupling regime.

## I. INTRODUCTION

Near to complete half a century since its first description, the Jaynes-Cummings Model (JCM) [1] still remains at the core of outstanding research. Most of the advances in new technologies and devices for quantum computing implementation based in solid state or atomic and optical systems made use of this model. Many tasks in quantum computing demand coherent superpositions and/or entanglement. These two intrinsically quantum ingredients are thus fundamental to disclose the quantum content of information. There has been a great effort to successfully and faithfully implement quantum computing building blocks in large scales. The simplest block is an information carrier capable of being in general superposition of two orthogonal states, also known as qubit [2]. Another fundamental block is the ability to perform general quantum state manipulations and measurements on these qubits [2].

The increasing degree of control that has been achieved in the level of single quantum systems has confirmed the predictions of fundamental models like the JCM and some of its generalizations [3]. The inclusion of a second radiation mode (bimodal cavities) is expected to bring out different quantum effects [4–7]. Such effects may concern quantum statistical properties of the fields, and also the dynamics with applications in the controlled generation of entanglement. The interaction of two-level atoms with two modes sustained by two weakly coupled cavities separated by a thin dielectric barrier is considered in [8], where the authors propose a generation scheme of atom-atom maximally entangled states. From the experimental side, the preparation of two modes of a superconducting cavity in a maximally entangled state has already been achieved [9]. Following this experimental achievement, generation schemes of important multipartite entangled states such as GHZ, W and cluster states [10, 11] have been suggested. In general, such schemes make use of a sudden change of the atomic transition frequency in order to induce resonant interaction with one mode or the other. This might introduce imperfections [12] that should be considered for the best agreement with the experiment.

In this paper, we study the dynamics of two non-degenerate bosonic modes interacting with $N$ identical two-level systems (qubits), aiming at the generation of important multipartite entangled states between the modes, qubits or modes and qubits (hybrid). An important feature of our proposal is that it does not make use of sudden frequency changes. By fixing the qubit frequency separation to be midway the natural frequencies of the modes [8], and by properly choosing the phases of the coupling constants, our schemes avoid the undesired effects introduced by suddenly switching the frequencies. Besides the generation of Bell, W and GHZ states, we also describe a one-step generation scheme of the linear cluster state among four qubits [13]. When comparing, for example, with cavity QED schemes, it does not make use of either rotation of qubits (Ramsey zones) or classical external fields, usual requirements in most schemes for entanglement generation in these setups [14, 15]. It is remarkable that the effective model obtained here (XY model) allows the generation of such a state, complementing the usual approach which employs the Ising model [13]. It is important to emphasize that all protocols for the generation of multipartite entangled states presented here are deterministic, i.e. no measurements are performed at all.

We organize the paper as follows. In Section II, we describe the model and its dynamics in the case of approximate resonance of the qubits with the bosonic modes considering: (i) not simultaneous interaction of qubits and modes, and (ii) simultaneous interaction of qubits and modes. We then discuss the generation of Bell and W states. In Section III, we derive the effective Hamiltonian of the qubits simultaneously interacting with the modes in the case of far off resonance, and we propose the generation of Bell, W, GHZ and cluster states in this configuration. In Section IV, we consider common experimental imperfections such as amplitude damping and finite time-control precision, analyzing the fidelity in the generation schemes and the violation of local realism for the cluster state. In Section V, we present physical setups where our ideas may be implemented. We summarize and conclude in Section VI. The ideas in this paper will be put in the general language of two-level systems (qubits) and bosonic modes. Consequently, the results presented here may be of interest for application in any particular quantum system described by JCM interactions. This includes circuit-QED systems [16] and trapped ions [17], just to name a few.

## II. GENERATION OF W STATES WITH BIMODAL JCM INTERACTIONS

In cavity-QED, phase differences between coupling constants of atoms and field modes are in general considered unimportant in many situations. However, special choices of these phases have been shown to avoid the destruction of mi-

---

croscopic correlations between the modes, important for the appearance of decoherence-free subspaces and robust states [18, 19]. Besides, it can not be underrated that these phases are of key importance for a plethora of quantum interference effects, responsible for many quantum phenomena [20]. In the context of trapped ions, adjustments of intensity and phase of the couplings constants have also been proved to be fundamental to the generation of cluster states when the ions are driven by red- and blue-sideband laser fields [21].

In order to explore such possibilities, we study a general system composed of $N$ two-level systems or qubits interacting with two bosonic modes. In the rotating wave approximation, the Hamiltonian for the two modes and qubit $k$ ($k = 1 \ldots N$), in the Schrödinger picture, reads ($\hbar = 1$)

$$H_k = \omega_A a^\dagger a + \omega_B b^\dagger b + \frac{\omega}{2}\sigma_k^z + \Omega_k^{(A)}(a^\dagger \sigma_k^- + a\sigma_k^+)$$
$$+ \Omega_k^{(B)}(b^\dagger \sigma_k^- + b\sigma_k^+), \quad (1)$$

where $\omega_A$ ($\omega_B$) is the frequency of mode $A$ ($B$), $a$ and $b$ ($a^\dagger$ and $b^\dagger$) are the photon annihilation (creation) operators of modes $A$ and $B$, $\omega$ is the qubit frequency and $\sigma_k^i$ ($i = x, y, z$), with $\sigma_k^\pm = (1/2)(\sigma_k^x \pm i\sigma_k^y)$, refer to the Pauli operators of qubit $k$. Here we consider the qubit coupling with mode $A$ to be real, positive, and constant i.e. $\Omega_k^{(A)} = \Omega$ for all $k$, while the qubit coupling with mode $B$ may also be negative, i.e. $\Omega_k^{(B)} = \Omega s_k$, where $s_k = \pm 1$.

We move to a frame rotating according to $e^{-iGt}$, with

$$G = \omega a^\dagger a + \omega b^\dagger b + \frac{\omega}{2}\sigma_k^z, \quad (2)$$

so that the Hamiltonian in this frame is given by

$$\tilde{H}_k = \Delta a^\dagger a - \Delta b^\dagger b + \Omega(a^\dagger \sigma_k^- + a\sigma_k^+) + \Omega s_k(b^\dagger \sigma_k^- + b\sigma_k^+), \quad (3)$$

where we have set the qubit frequency midway the modes frequencies separation, i.e. the detuning is given by $\Delta = \omega_A - \omega = \omega - \omega_B$, with $\omega_A > \omega_B$.

### A. One qubit per time

In cavity-QED, this situation is precisely the one where two-level atoms are sent one by one through a bimodal cavity in such a way that only one atom is found inside the cavity per time. We then let the first qubit interact with the modes. If the system state at $t = 0$ is $|0_A\rangle|0_B\rangle|\uparrow_1\rangle \equiv |00\uparrow\rangle$, where the modes are in the vacuum state and the qubit 1 is in the spin up state[1], we may restrict the space state to the subspace spanned by the basis $\{|10\downarrow\rangle, |01\downarrow\rangle, |00\uparrow\rangle\}$. After the interaction of qubit 1 with the modes (time $t$), the system state will be given by

$$|\psi(t)\rangle = c_1(t)|10\downarrow\rangle + c_2(t)|01\downarrow\rangle + c_3(t)|00\uparrow\rangle, \quad (4)$$

---

[1] We borrowed the usual notation of spin−1/2 particles when referring to the states of the qubits.

with coefficients

$$c_1(t) = -\frac{\Omega}{\tilde{\Omega}^2}[\Delta(1 - \cos\tilde{\Omega}t) + i\tilde{\Omega}\sin\tilde{\Omega}t], \quad (5)$$

$$c_2(t) = -s_1 c_1^*(t), \quad (6)$$

$$c_3(t) = \frac{1}{\tilde{\Omega}^2}(\Delta^2 + 2\Omega^2 \cos\tilde{\Omega}t), \quad (7)$$

where $\tilde{\Omega} = \sqrt{\Delta^2 + 2\Omega^2}$ is the Rabi frequency. The probability of finding this qubit in the spin up state is

$$P_{\uparrow_1}(t) = |c_3(t)|^2 = \frac{1}{\tilde{\Omega}^4}(\Delta^2 + 2\Omega^2 \cos\tilde{\Omega}t)^2. \quad (8)$$

This quantity plays an important role in the generation protocols described in this section. We may use (8) to obtain the shortest time needed to achieve a desired value of $P_{\uparrow_1}$

$$t = \frac{1}{\tilde{\Omega}} \arccos \frac{\tilde{\Omega}^2 \sqrt{P_{\uparrow_1}} - \Delta^2}{2\Omega^2}, \quad (9)$$

from where we can see that

$$0 \leq \frac{\Delta}{\Omega} \leq \sqrt{2}\frac{\sqrt{1 + \sqrt{P_{\uparrow_1}}}}{\sqrt{1 - \sqrt{P_{\uparrow_1}}}}, \quad (10)$$

because the domain of the arccos function is finite. Thus, the qubit 1 may transfer its whole excitation energy to the field modes ($P_{\uparrow_1} = 0$) whenever $0 \leq \Delta \leq \sqrt{2}\Omega$, in accordance with [8]. In this case, since $|c_1(t)|^2 = |c_2(t)|^2$, the non-degenerate modes will end up in the Bell state

$$|\Psi^+\rangle = \frac{1}{\sqrt{2}}(|10\rangle + |01\rangle), \quad (11)$$

up to local unitary (l.u.) operations. It is important to remark that no measurements were needed here because the qubit disentangles with the modes after such $t$. In the cavity-QED language, for instance, in order to generate this state for two field modes, previous schemes [22, 23] generally made use of either two cavities (making the experimental control and costs more demanding) or a sudden change of the atomic frequency also in a bimodal setup (which can introduce errors and decrease the fidelity of the generation protocol). However, if one is interested in entangling distant cavities for teleportation purposes [24], a two cavity setup is of course more appropriate than a single bimodal cavity.

Conversely, if the first qubit ends up with $P_{\uparrow_1} > 0$, it will be entangled with the modes. For instance, consider the case in which $P_{\uparrow_1} = 1/3$. This may be achieved by choosing $0 \leq \Delta \leq (1 + \sqrt{3})\Omega$ and adjusting $t$ through (9). In this case, the system will end up in the state (l.u.)

$$|W_3\rangle = \frac{1}{\sqrt{3}}(|10\downarrow\rangle + |01\downarrow\rangle + |00\uparrow\rangle), \quad (12)$$

which is a hybrid W state among these three subsystems. Another example is the choice $P_{\uparrow_1} = 1/2$, which demands $0 \leq \Delta \leq (2 + \sqrt{2})\Omega$. Again, by adjusting $t$ with the help of (9), a hybrid entangled state (l.u.)

$$|W_T\rangle = \frac{1}{2}(|10\downarrow\rangle + |01\downarrow\rangle + \sqrt{2}|00\uparrow\rangle), \quad (13)$$

is generated. This is a special instance of a generalized W state, and it finds application in quantum teleportation [25].

In what concerns the generation of entangled states, the role of the sign of coupling constant seems to be irrelevant for the one qubit setup. However, when a second qubit subsequently interacts with the modes, the choice of different constant coupling phases will lead to substantially different dynamics. For completeness, we will also consider a delay time $t_d$ during which the subsystems follow free evolution [8]. If the interaction of qubit 1 with the modes takes a time $t_1$, and the interaction of qubit 2 (initially prepared in the spin down state $|\downarrow\rangle$) with the modes takes a time $t_2$, the final global state can be written as

$$|\psi(t_1, t_d, t_2)\rangle = \alpha(t_1, t_d, t_2)|10\downarrow\downarrow\rangle + \beta(t_1, t_d, t_2)|01\downarrow\downarrow\rangle \\ + \gamma(t_1)|00\uparrow\downarrow\rangle + \delta(t_1, t_d, t_2)|00\downarrow\uparrow\rangle, \quad (14)$$

where we point out the independence of $\gamma(t_1) \equiv c_3(t_1)$ on subsequent times. The other coefficients are given by

$$\alpha(t_1, t_d, t_2) = c_1(t_1)a(t_2)e^{-i\Delta t_d} + c_2(t_1)b(t_2)e^{i\Delta t_d}, \\ \beta(t_1, t_d, t_2) = c_1(t_1)b(t_2)e^{-i\Delta t_d} + c_2(t_1)a^*(t_2)e^{i\Delta t_d}, \\ \delta(t_1, t_d, t_2) = c_1(t_1)c_1(t_2)e^{-i\Delta t_d} + c_2(t_1)c_2(t_2)e^{i\Delta t_d}, \quad (15)$$

with $c_1(t_j)$ and $c_2(t_j)$ obtained from (5) and (6), respectively, and

$$a(t_2) = \frac{1}{\tilde{\Omega}^2}[\Omega^2 + (\Omega^2 + \Delta^2)\cos\tilde{\Omega}t_2 - i\Delta\tilde{\Omega}\sin\tilde{\Omega}t_2], \quad (16)$$

$$b(t_2) = -s_2 \frac{\Omega^2}{\tilde{\Omega}^2}(1 - \cos\tilde{\Omega}t_2). \quad (17)$$

We now provide a link between the sign of the coupling constants and the delay time. We will show that the choice of the relative sign of coupling constants may be chosen to mimic the delay time under special conditions. In order to see that, one has to realize that the probability of finding the qubit 2 in the spin up state $P_{\uparrow_2}(t_1, t_d, t_2) = |\delta(t_1, t_d, t_2)|^2$ can be written, for the choice $t_d = j\pi/(2\Delta)$ ($j = 0, 1, 2, \ldots$), as

$$P_{\uparrow_2}(t_1, t_2) = 2\{|c_1(t_1)|^2|c_1(t_2)|^2 \\ + (-1)^j s_1 s_2 \text{Re}[c_1^2(t_1)c_1^2(t_2)]\}. \quad (18)$$

Therefore, the choice of $s_1 = -s_2$ together with no delay time ($j = 0$) is equivalent to $s_1 = s_2$ with $j$ odd. The possibility of eliminating the delay time through careful choices of the phases of the coupling constants may be useful for diminishing decoherence damage (reducing overall time). This is a quantum interference effect.

The pursuit of generation of $W$ states involving more qubits in this scenario may not be useful since the number of steps in this generation scheme grows with the number of qubits, and decoherence is expected to decrease the state fidelity if many steps are required. This motivates the search for schemes in which the generation of $N$-qubit entangled states does not scale with $N$. The rest of the paper is dedicated to such schemes. Indeed, it turns out that the next scheme does not scale with $N$ (one step generation), and the time taken to complete the protocol decreases with $N$.

### B. $N$ qubits at a time

The system Hamiltonian for $N$ qubits interacting with two bosonic modes in the rotating wave approximation reads

$$H = \omega_A a^\dagger a + \omega_B b^\dagger b + \frac{\omega}{2}\sum_{k=1}^{N}\sigma_k^z + \sum_{k=1}^{N}\Omega_k^{(A)}(a^\dagger\sigma_k^- + a\sigma_k^+) \\ + \sum_{k=1}^{N}\Omega_k^{(B)}(b^\dagger\sigma_k^- + b\sigma_k^+). \quad (19)$$

Please note that we are disregarding qubit-qubit interactions. In cavity-QED setups, for example, this corresponds to maintain the atoms sufficiently apart so that direct dipole-dipole interactions may be neglected.

We now prepare the system in the state $|00\rangle\sigma_1^+|\downarrow\rangle^{\otimes N}$, where each mode is in the vacuum state, qubit 1 in the spin up state, and any other qubit in the spin down state. We will consider the case where all the coupling constants have the same sign. Thus, similar to (2) and (3), we now move to a frame rotating according to

$$G_N = \omega a^\dagger a + \omega b^\dagger b + \frac{\omega}{2}\sum_{k=1}^{N}\sigma_k^z, \quad (20)$$

and the evolved state at time $t$ will be [26, 27]

$$|\psi_N(t)\rangle = a_N(t)|10\rangle|\downarrow\rangle^{\otimes N} + b_N(t)|01\rangle|\downarrow\rangle^{\otimes N} \\ + \sum_{k=1}^{N} c_N^{(k)}(t)|00\rangle\sigma_k^+|\downarrow\rangle^{\otimes N}, \quad (21)$$

with the coefficients

$$a_N(t) = -\frac{\Omega}{\tilde{\Omega}_N^2}[\Delta(1 - \cos\tilde{\Omega}_N t) + i\tilde{\Omega}_N \sin\tilde{\Omega}_N t], \quad (22)$$

$$b_N(t) = \frac{\Omega}{\tilde{\Omega}_N^2}[\Delta(1 - \cos\tilde{\Omega}_N t) - i\tilde{\Omega}_N \sin\tilde{\Omega}_N t], \quad (23)$$

$$c_N^{(1)}(t) = 1 - \frac{2\Omega^2}{\tilde{\Omega}_N^2}(1 - \cos\tilde{\Omega}_N t), \quad (24)$$

$$c_N^{(k\neq 1)}(t) = c_N^{(1)}(t) - 1, \quad (25)$$

where $\tilde{\Omega}_N = \sqrt{\Delta^2 + 2N\Omega^2}$. Of course, for $N = 1$ we recover the results of previous analysis (observe that the last equation above does not apply in this case). For $N > 1$, it follows from the last two equations that

$$t = \frac{1}{\tilde{\Omega}_N}\arccos\frac{\tilde{\Omega}_N^2(\sqrt{P_{\uparrow_1}^{(N)}} - 1) + 2\Omega^2}{2\Omega^2} \\ = \frac{1}{\tilde{\Omega}_N}\arccos\frac{\tilde{\Omega}_N^2\sqrt{P_{\uparrow_k}^{(N)}} + 2\Omega^2}{2\Omega^2}, \quad (26)$$

is the shortest time needed to achieve a given probability $P_{\uparrow_{1(k)}}^{(N)}$ of finding qubit 1 ($k \neq 1$) in the spin up state. Also, from the arguments of the arccos functions above, these two probabilities are related by

$$\sqrt{P_{\uparrow_1}^{(N)}} - 1 = \sqrt{P_{\uparrow_k}^{(N)}}. \quad (27)$$

As we are interested in the generation of hybrid (prototype) $W_{N+2}$ ($W_N$) states, we must impose $P_{\uparrow_1}^{(N)} = P_{\uparrow_k}^{(N)} = 1/(N+2)$ ($P_{\uparrow_1}^{(N)} = P_{\uparrow_k}^{(N)} = 1/N$) which is only fulfilled for $N = 2$ ($N = 4$), see (27). Thus, for $N > 1$, this initial preparation *only* allows for the generation of hybrid or prototype W states involving either two modes and two qubits or four qubits (modes factoring out). Such a limitation may also be observed in similar situations where only one excitation is present in the system [28, 29]. Here, the hybrid (prototype) $W_4$ state may be generated for a detuning satisfying $0 \leq \Delta \leq 2\Omega$ ($\Delta = 0$).

The limitation on $N$ for generation of hybrid (prototype) W states may be avoided if we follow a *prior* step: we let an auxiliary spin up qubit (frequency $\omega_0$) interact with the modes (initially in vacuum states) transferring its whole energy to the modes [(9) and (10) with $P_{\uparrow_0}(t) = 0$]. This is the exact protocol described in the last section. Conveniently parameterizing the resulting modes state as $|\psi_c\rangle = p|10\rangle - s_0 p^*|01\rangle$, where $s_0 = -1$ is the coupling sign of the auxiliary qubit with mode $B$ and

$$p = -\frac{1}{2}(\Delta_0/\Omega + i\sqrt{2 - \Delta_0^2/\Omega^2}), \tag{28}$$

with $\Delta_0 = \omega_A - \omega_0 = \omega_0 - \omega_B$, it follows that $|p|^2 = 1/2$ whenever $0 \leq \Delta_0 \leq \sqrt{2}\Omega$, leaving the fields in $|\psi_c\rangle$ which is l.u. to the Bell state (11). The auxiliary qubit at the end of this step is in the spin down state and is disregarded thereafter.

We now let $N$ qubits, each of them initially in the spin down state, simultaneously interact with the *entangled* modes. We again consider the case where all the coupling signs $s_k$ are positive. The evolved state of the system after an interaction time $t$ is still given by (21), except for the coefficients that now read

$$a_N(t) = \frac{1}{\tilde{\Omega}_N^2}\{(p - p^*)N\Omega^2 + [p\Delta^2 + (p + p^*)N\Omega^2]$$
$$\times \cos\tilde{\Omega}_N t - ip\Delta\tilde{\Omega}_N \sin\tilde{\Omega}_N t\}, \tag{29}$$

$$b_N(t) = \frac{1}{\tilde{\Omega}_N^2}\{(p^* - p)N\Omega^2 + [p^*\Delta^2 + (p + p^*)N\Omega^2]$$
$$\times \cos\tilde{\Omega}_N t + ip^*\Delta\tilde{\Omega}_N \sin\tilde{\Omega}_N t\}, \tag{30}$$

$$c_N^{(k)}(t) = \frac{\Omega}{\tilde{\Omega}_N^2}[(p^* - p)\Delta(1 - \cos\tilde{\Omega}_N t) - i(p + p^*)\tilde{\Omega}_N$$
$$\times \sin\tilde{\Omega}_N t]. \tag{31}$$

Please note that there is no differentiation between the amplitudes of probability (referred to the spin up state) for the qubit 1 and the others. This fact will give us freedom to generate W states with any number of qubits. In the following we consider $p$ real ($\Delta_0 = \sqrt{2}\Omega$) or imaginary ($\Delta_0 = 0$). For a $p$ real, it follows that

$$t = \frac{1}{\tilde{\Omega}_N} \arcsin \frac{\tilde{\Omega}_N \sqrt{P_{\uparrow_k}^{(N)}}}{\sqrt{2}\Omega}, \tag{32}$$

is the shortest time needed to achieve a given probability $P_{\uparrow_k}^{(N)}(t)$ of finding qubit $k$ in the spin up state, with a detuning in the interval

$$0 \leq \frac{\Delta}{\Omega} \leq \sqrt{2}\sqrt{\frac{1}{P_{\uparrow_k}^{(N)}} - N}. \tag{33}$$

While for a $p$ imaginary we have

$$t = \frac{1}{\tilde{\Omega}_N} \arccos \frac{\sqrt{2}\Omega\Delta - \tilde{\Omega}_N^2 \sqrt{P_{\uparrow_k}^{(N)}}}{\sqrt{2}\Omega\Delta}, \tag{34}$$

with the detuning now in the interval

$$\sqrt{2}\frac{(1 - \sqrt{1 - NP_{\uparrow_k}^{(N)}})}{\sqrt{P_{\uparrow_k}^{(N)}}} \leq \frac{\Delta}{\Omega} \leq \sqrt{2}\frac{(1 + \sqrt{1 - NP_{\uparrow_k}^{(N)}})}{\sqrt{P_{\uparrow_k}^{(N)}}}. \tag{35}$$

In order to generate $W_{N+2}$ ($W_N$) states, we must impose $P_{\uparrow_k}^{(N)} = 1/(N+2)$ ($P_{\uparrow_k}^{(N)} = 1/N$) in the above equations. This implies that hybrid (prototype) W states occur if the detuning is such that $0 \leq \Delta \leq 2\Omega$ ($\Delta = 0$) for $p$ real or $(\sqrt{2}\sqrt{N+2} - 2)\Omega \leq \Delta \leq (\sqrt{2}\sqrt{N+2} + 2)\Omega$ ($\Delta = \sqrt{2N}$) for $p$ imaginary. Careful analysis of (32) and (34) reveals that the time required to generate $W_{N+2}$ or $W_N$ states decreases with $N$. A similar result is presented in [30]. We emphasize that due to its importance in the multipartite entanglement theory [31], many different schemes for generation of W states have been previously proposed [32–40], and their experimental realization has also been reported [41–45].

## III. GENERATION OF W, GHZ, AND CLUSTER STATES WITH BIMODAL DISPERSIVE INTERACTIONS

It is a well known fact that a single bosonic mode may mediate interactions between two previously non-interacting qubits [46]. Consider for example that the mode frequency is sufficiently apart from the qubit frequencies (taken to be all equal), the so-called dispersive limit. In the two-qubit case, the mode in the vacuum state then induces a dipole-dipole interaction between the two qubits ($XY$ interaction), allowing the generation of a Bell state at certain times of interaction [46]. In the following, we consider the dynamics far from resonance of two modes simultaneously interacting with $N$ qubits, and explore the role played by the phases of the coupling constants.

The simultaneous interaction of $N$ qubits with two non-degenerate modes is described in the interaction picture by the Hamiltonian

$$H_I(t) = \Omega \sum_{k=1}^{N}(a^\dagger \sigma_k^- e^{i\Delta t} + a\sigma_k^+ e^{-i\Delta t})$$
$$+ \Omega \sum_{k=1}^{N} s_k(b^\dagger \sigma_k^- e^{-i\Delta t} + b\sigma_k^+ e^{i\Delta t}), \tag{36}$$

where we have again set the qubit frequencies ($\omega$) midway the modes frequency separation, so that $\Delta = \omega_A - \omega = \omega - \omega_B$, with $\omega_A > \omega_B$.

The dispersive limit is obtained when the detuning $\Delta$ is sufficiently larger than the coupling constant $\Omega$. Following the method described in [47], we obtain from (36) the effective Hamiltonian

$$H_{\text{eff}} = -\lambda(a^\dagger a - b^\dagger b)\sum_{k=1}^{N}\sigma_k^z - \lambda\sum_{j,k}^{N}(1 - s_j s_k)\sigma_j^+ \sigma_k^-, \tag{37}$$

where $\lambda = \Delta^2/\Omega$ is the effective qubit-qubit coupling constant induced by the modes.

## A. W states

For $N = 1$, the above Hamiltonian has already been considered for the generation of multidimensional entangled coherent states of two cavity modes [48]. The first term of the above equation describes the Stark-shifts. These induced shifts may cancel each other if the modes are prepared in Fock states with equal number of photons (or simply in the vacuum). The second term (dipole-dipole interactions induced by the modes) may be totally inhibited if $s_k = 1$ for all $k$. This occurs because the interaction between qubits $j$ and $k$ induced by the mode $A$ cancels the interaction induced by the mode $B$. Clearly, this effect is absent when only one mode interacts with the qubits. We may then state a general rule that given two qubits with $s_j$ and $s_k$, the induced dipole-dipole interaction between them will be inhibited whenever $s_j = s_k$. The contrary is also true, i.e. their induced dipole-dipole interaction will be doubled provided $s_j = -s_k$.

We still consider cancellation of the Stark-shifts in (37). We then consider that only one of the $N$ qubits is initially in the spin up state, say the first qubit, and the others are in the fundamental state, so that the initial state of the qubits is $\sigma_1^+|\downarrow\rangle^{\otimes N}$. If we set $s_1 = -1$ and $s_k = 1$, for $k > 1$, the state of the $N$ qubits after *simultaneously* interacting with modes will be (at a time $t$)

$$|\psi_N(t)\rangle = \sum_{k=1}^{N} c_N^{(k)}(t)\sigma_k^+|\downarrow\rangle^{\otimes N}, \quad (38)$$

with coefficients

$$c_N^{(1)}(t) = \cos(2\sqrt{N-1}\lambda t), \quad (39)$$

$$c_N^{(k\neq 1)}(t) = i\frac{\sin(2\sqrt{N-1}\lambda t)}{\sqrt{N-1}}. \quad (40)$$

From this follows that

$$t = \frac{\arccos\sqrt{P_{\uparrow_1}^{(N)}}}{2\sqrt{N-1}\lambda}, \quad (41)$$

is the shortest time needed to achieve a given $P_{\uparrow_1}^{(N)}$, i.e. the probability of finding the qubit 1 in the spin up state. It is interesting to notice that now

$$1 - P_{\uparrow_1}^{(N)} = (N-1)P_{\uparrow_k}^{(N)}, \quad (42)$$

so that the qubits after a time $t$ are left in the state

$$|\psi_N(t)\rangle = \sqrt{P_{\uparrow_1}^{(N)}}\sigma_1^+|\downarrow\rangle^{\otimes N} + i\sqrt{\frac{1-P_{\uparrow_1}^{(N)}}{N-1}}\sum_{k=2}^{N}\sigma_k^+|\downarrow\rangle^{\otimes N}. \quad (43)$$

This means that a wide class of W states may be generated using the present scheme. For example, if $N = 3$, we may generate the usual (prototype) $W_3$ or its variant (suited for teleportation) just by setting $P_{\uparrow_1}^{(3)} = 1/3$ or $P_{\uparrow_1}^{(3)} = 1/2$, respectively. As in the last section, analysis of (41) also reveals the decreasing of interaction time with the number of qubits.

## B. GHZ states

This configuration is also useful for generating more kinds of multipartite entangled states when we leave the one-excitation sector of the system Hamiltonian. For example, if the qubits are prepared in the state $|+_1\rangle|+_2\rangle|+_3\rangle \equiv |+++\rangle$, with $|+_j\rangle = (1/\sqrt{2})(|\uparrow_j\rangle+|\downarrow_j\rangle)$, the evolved state after a time $t$ will be

$$|\psi(t)\rangle = \frac{1}{2\sqrt{2}}[|\uparrow\uparrow\uparrow\rangle + (\mu+i\nu)|\uparrow\uparrow\downarrow\rangle + (\mu+i\nu)|\uparrow\downarrow\uparrow\rangle$$
$$+ (\mu+2i\nu)|\uparrow\downarrow\downarrow\rangle + (\mu+2i\nu)|\downarrow\uparrow\uparrow\rangle$$
$$+ (\mu+i\nu)|\downarrow\uparrow\downarrow\rangle + (\mu+i\nu)|\downarrow\downarrow\uparrow\rangle + |\downarrow\downarrow\downarrow\rangle], \quad (44)$$

where $\mu = \cos 2\sqrt{2}\lambda t$ and $\nu = (1/\sqrt{2})\sin 2\sqrt{2}\lambda t$. Hence, if we let the qubits interact with the modes for $\lambda t = \pi/(2\sqrt{2})$, the following state will be generated

$$|\psi\rangle = \frac{1}{2\sqrt{2}}(|\uparrow\uparrow\uparrow\rangle - |\uparrow\uparrow\downarrow\rangle - |\uparrow\downarrow\uparrow\rangle - |\uparrow\downarrow\downarrow\rangle$$
$$- |\downarrow\uparrow\uparrow\rangle - |\downarrow\uparrow\downarrow\rangle - |\downarrow\downarrow\uparrow\rangle + |\downarrow\downarrow\downarrow\rangle), \quad (45)$$

which is l.u. to the GHZ state, e.g. $VU|\psi\rangle = (1/\sqrt{2})(|\uparrow\uparrow\uparrow\rangle + |\downarrow\downarrow\downarrow\rangle)$, where $U = (-i\sigma_1^x)^{1/2} \otimes (i\sigma_2^z)^{1/2} \otimes (i\sigma_3^z)^{1/2}$ and $V = (-\sqrt{i})\sigma_1^z \otimes \mathcal{H}_2 \otimes \mathcal{H}_3$, with $\mathcal{H}_i$ representing the Hadamard operation on qubit $i$. It is known that $U$ belongs to a set of local Clifford unitaries mapping the whole equivalence class associated with a given graph state [49–51]. The GHZ state has an important role in the study of fundamentals of quantum mechanics. Greenberger, Horne and Zeilinger (GHZ) have shown that the predictions of local realistic theories and quantum mechanics are completely different using this state [52]. Other theoretical proposals for generation of GHZ states have been previously published [53–56] as well as its experimental realization [57–61].

## C. Cluster states

Now consider $N = 4$, $s_1 = s_2 = -1$ and $s_3 = s_4 = 1$. In this case, Hamiltonian (37) reads (modes in the vacuum)

$$H_{\text{eff}} = -2\lambda[(\sigma_1^+\sigma_3^- + \sigma_1^-\sigma_3^+) + (\sigma_1^+\sigma_4^- + \sigma_1^-\sigma_4^+)$$
$$+ (\sigma_2^+\sigma_3^- + \sigma_2^-\sigma_3^+) + (\sigma_2^+\sigma_4^- + \sigma_2^-\sigma_4^+)]. \quad (46)$$

If the initial state of the qubits is $|\uparrow\downarrow\uparrow\downarrow\rangle$, the evolved state at $\lambda t = \pi/(4\sqrt{2})$ will be

$$|\psi\rangle = \frac{1}{2}(|\uparrow\downarrow\uparrow\downarrow\rangle - |\uparrow\downarrow\downarrow\uparrow\rangle - |\downarrow\uparrow\uparrow\downarrow\rangle - |\downarrow\uparrow\downarrow\uparrow\rangle), \quad (47)$$

which is l.u. to the usual form of a linear cluster state [13], e.g. $-\sigma_1^x \otimes \mathbb{1}_2 \otimes \sigma_3^x \otimes \mathbb{1}_4|\psi\rangle = (1/2)(|\uparrow\uparrow\uparrow\uparrow\rangle + |\uparrow\uparrow\downarrow\downarrow\rangle + |\downarrow\downarrow\uparrow\uparrow\rangle - |\downarrow\downarrow\downarrow\downarrow\rangle)$. Multidimensional cluster states form the basis of what is called one-way quantum computing [62]. By initializing the system in these highly entangled states, and by properly choosing local measurements on the qubits (plus classical feed-forward), it is possible to achieve a universal model for quantum computing [62]. Previous schemes for generation of these states have been proposed [63–67], and experimental realization has also been reported [68–71].

## IV. GENERATION SCHEMES IN THE PRESENCE OF IMPERFECTIONS

In this section we analyze the effects of usual imperfections that may arise in a realistic setup. Whereas the preceding results were put in the general language of bosonic modes and qubits, the sources of imperfections we are going to discuss are well suited for the cavity QED context, where two-level atoms (qubits) can interact with bosonic modes sustained by a cavity. We separate the analysis of imperfections in three subsections. In Section IV A, we investigate the effect of dissipative environments in the generation schemes of previous sections. In Section IV B, we estimate the error in the state preparation arising from the intrinsic uncontrollability of the time of flight of atoms. Both sources of imperfections will be analyzed in terms of the fidelity $F = \sqrt{\langle\psi|\rho|\psi\rangle}$ [2], where $|\psi\rangle$ is the ideally generated (target) state and $\rho$ is the result in the presence of imperfections. In Section IV C, we show the possibility of violation of local realism considering our generation protocol of cluster states in the presence of such imperfections.

### A. Effects of dissipative environments

In real experiments, one usually faces the effects of many decoherence channels. For the system considered here, the most relevant damages to the dynamics come from photon leakage from cavity mirrors and fluorescent radiation from the atoms (spontaneous emission). The usual formalism to treat irreversible decaying processes considers the coupling of the system to large reservoirs, generally modeled as baths of harmonic oscillators [72]. Here we consider zero temperature local reservoirs for each subsystem.

We start by considering dissipation in the generation schemes of Section II. In the rotating-wave and Born-Markov approximations, the density operator for the system obeys the following master equation [72]

$$\dot{\rho}(t) = -i[\tilde{H}_k, \rho(t)] + \mathcal{L}_A\rho(t) + \mathcal{L}_B\rho(t) + \sum_{k=1}^{N}\mathcal{L}_k\rho(t), \quad (48)$$

where $\tilde{H}_k$ is given by (3),

$$\mathcal{L}_A\rho(t) = \frac{\kappa_A}{2}([a,\rho(t)a^\dagger] + [a\rho(t),a^\dagger]), \quad (49)$$

$$\mathcal{L}_B\rho(t) = \frac{\kappa_B}{2}([b,\rho(t)b^\dagger] + [b\rho(t),b^\dagger]), \quad (50)$$

$$\mathcal{L}_k\rho(t) = \frac{\gamma}{2}([\sigma_k^-,\rho(t)\sigma_k^+] + [\sigma_k^-\rho(t),\sigma_k^-]), \quad (51)$$

are the Lindblad operators acting on the density operator $\rho(t)$, $\kappa_A$, $\kappa_B$, and $\gamma$ are the decay rates for cavity mode $A$, $B$ and the atoms, respectively.

In Figure 1, we plot the fidelity between $\rho$ and $|\psi\rangle$ as a function of $\chi/\Omega$ for three different target states. In these calculations, $\rho$ is also considered at an interaction time given by (9) with $\Delta = \sqrt{2}\Omega$. The parameter $\chi$ is used to embody different configurations for the decay rates. These plots show that the worst situation corresponds to that of equal dissipation rates (cavity modes and atoms). It is noteworthy that the generation protocol aiming at the $W_T$ state (13) is more robust to dissipative effects when compared to the ones for $W_3$ state (12) or the cavity Bell state (11).

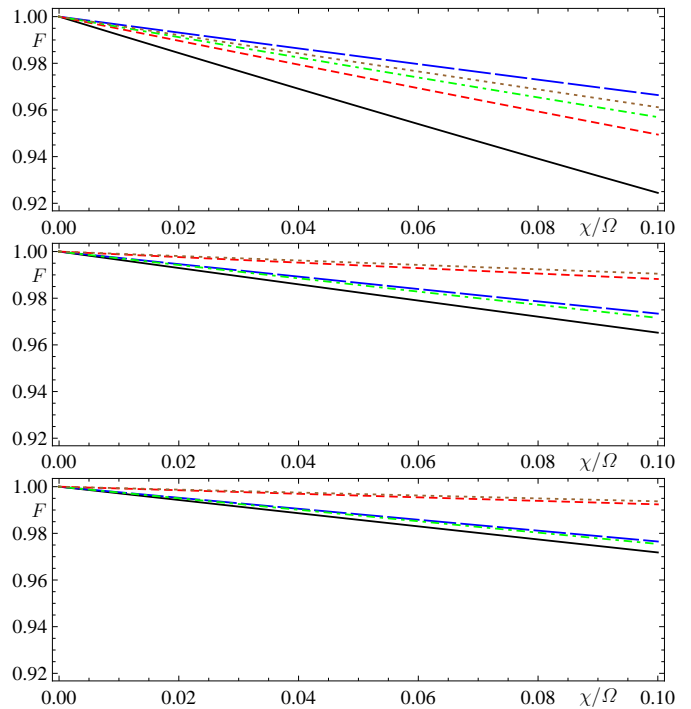

FIG. 1. (Color online) Fidelity for the cavity Bell state (top), hybrid $W_3$ state (middle), and hybrid $W_T$ state (bottom). The configurations are $\chi = \kappa_A = \kappa_B = \gamma$ (black solid line), $\chi = 0.1\kappa_A = 0.1\kappa_B = \gamma$ (blue large-dashed line), $\chi = \kappa_A = \kappa_B = 0.1\gamma$ (red dashed line), $\chi = 0.1\kappa_A = 0.5\kappa_B = \gamma$ or $\chi = 0.5\kappa_A = 0.1\kappa_B = \gamma$ (green dot-dashed line), and $\chi = \kappa_A = 0.5\kappa_B = 0.1\gamma$ or $\chi = 0.5\kappa_A = \kappa_B = 0.1\gamma$ (brown dotted line).

In order to compare the quality of the protocols for generation of W and GHZ states under dissipation, we move now to the cases treated in Section III. In these cases, the master equation (48) is still valid with the only modification being the replacement of $\tilde{H}_k$ for $H_{\text{eff}}$, the latter given by (37). Once the modes are still prepared in vacuum states, we may still disregard the Stark-shifts appearing in (37) as well as any losses due to cavity damping. It can now be seen from Figure 2 that the W states protocols are again more robust in the presence of losses, now compared to the protocol for generating the GHZ$_3$ state. Notably, the generation scheme for $W_T$ is still the most robust.

The analysis of imperfections for the cluster state is pre-sented in a separate section later, where we discuss applications in experiments for testing violation of local realism.

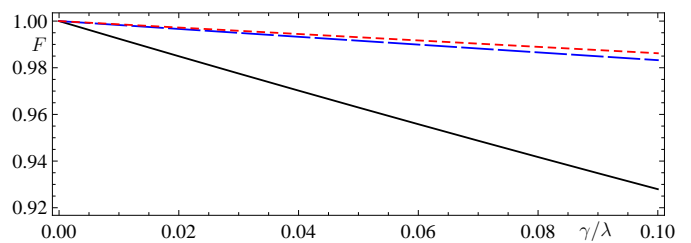

FIG. 2. (Color online) Fidelity for the GHZ$_3$ state (black solid line), $W_3$ state (blue large-dashed line), and $W_T$ state (red dashed line).

### B. Errors in the control of atomic time of flight

In majority of the generation schemes presented in this article, two or more atoms are required to simultaneously interact with the cavity modes. In realistic setups, however, it might happen that the atoms will not enter the cavity exactly at the same time, consequently inducing an error in the target state generation. In the case involving only two atoms, the inclusion of such error source is a simple matter as discussed, for instance, in reference [46]. The approach consists of supposing one of the atoms reaching the cavity in advance or in delay compared to the alleged initial interaction time.

We need to generalize this approach in order to analyze this source of error in our protocols. We do this by choosing the time each atom enters the cavity from a normal (Gaussian) distribution centered at $t = 0$ (the ideal time for entering the cavity). The uncontrollability of the time of flights translates then as finite standard deviations for the normal distribution. In Figure 3, each point in the plotted fidelities corresponds to an average over many realizations, each one having the same standard deviation measured as a fraction (percentage) of $1/\lambda$.

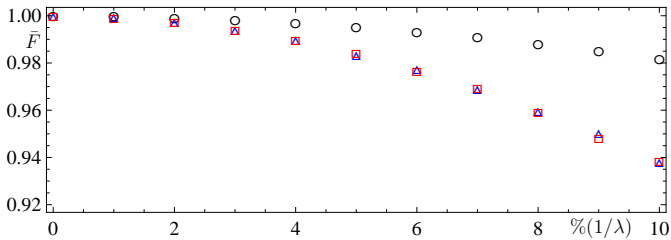

FIG. 3. (Color online) Average value of fidelity as a function of the fluctuation of the atomic times of flight measured as a fraction (percentage) of $1/\lambda$ for GHZ$_3$ (black circles), W$_3$ (blue triangles), and W$_T$ (red squares). Here, 3,000 repetitions have been considered to evaluate the averaged fidelity.

It is a remarkable fact that the generation of W$_T$ and W$_3$ are now affected in almost the same way by moderate fluctuations in the interaction times. Also, the GHZ$_3$ state protocol, which was previously shown to be more sensitive to energy relaxation, turns out now to be more robust than the other protocols for errors in the time control.

### C. Cluster state nonlocality under imperfections

In reference [73], Scarani, Acín, Schenck and Aspelmeyer introduced a nonlocality test (hereafter mnemonically referred to as SASA inequality) suitable for cluster states. The SASA operator is given by [73]

$$\mathcal{B} = \sigma_1^x \otimes \mathbb{1}_2 \otimes \sigma_3^x \otimes \sigma_4^z + \sigma_1^x \otimes \mathbb{1}_2 \otimes \sigma_3^y \otimes \sigma_4^y$$
$$+ \sigma_1^z \otimes \sigma_2^y \otimes \sigma_3^y \otimes \sigma_4^z - \sigma_1^z \otimes \sigma_2^y \otimes \sigma_3^x \otimes \sigma_4^y, \quad (52)$$

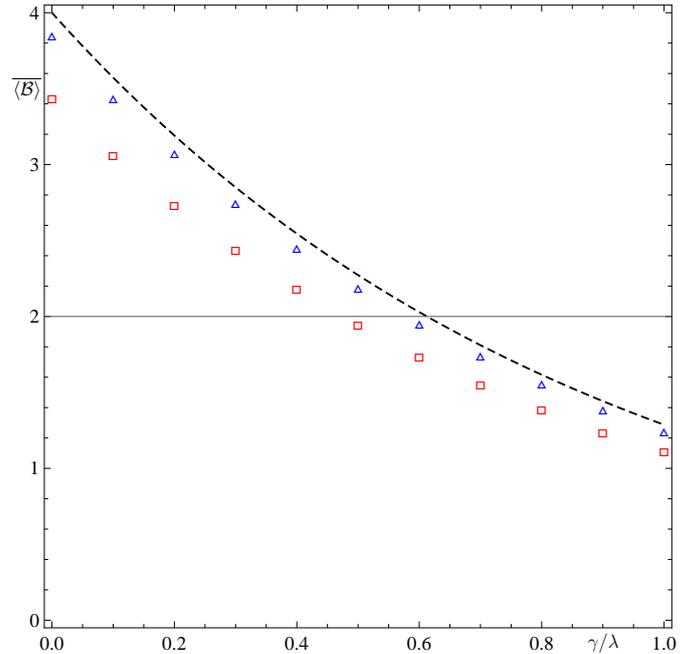

FIG. 4. (Color online) Average value of the SASA operator as a function of $\gamma/\lambda$ for different levels of precision in the control of the time of flight of the atoms, measured as percentages of $1/\lambda$, for $0\%$ (black dashed line), $5\%$ (blue triangles), and $10\%$ (red squares). Here, 3,000 repetitions have been considered to evaluate the average $\overline{\langle \mathcal{B} \rangle}$.

and any local realistic theory will lead to $\langle \mathcal{B} \rangle \leq 2$. For the cluster state

$$|\phi_4\rangle = \frac{1}{2}|+\rangle|\uparrow\rangle|+\rangle|\uparrow\rangle + \frac{1}{2}|+\rangle|\uparrow\rangle|-\rangle|\downarrow\rangle$$
$$+ \frac{1}{2}|-\rangle|\downarrow\rangle|-\rangle|\uparrow\rangle + \frac{1}{2}|-\rangle|\downarrow\rangle|+\rangle|\downarrow\rangle, \quad (53)$$

where $|\pm\rangle = (|\uparrow\rangle \pm |\downarrow\rangle)/\sqrt{2}$, it is found maximal violation, i.e. $\langle \mathcal{B} \rangle = 4$. The experimental verification of violation of the SASA inequality by linear cluster states l.u. to (53) has been demonstrated [74, 75].

In order to evaluate the expectation value of the SASA operator in form (52), using the cluster state $|\psi\rangle$ generated in our protocol, all one have to realize is that $T|\psi\rangle = |\phi_4\rangle$ with

$$T = -\mathcal{H}_1 \sigma_1^x \otimes \mathbb{1}_2 \otimes \sigma_3^x \otimes \mathcal{H}_4. \quad (54)$$

Following the same guidelines explained in the last subsections, we now analyze the effect of dissipation and time of flight fluctuations on the violation of the SASA inequality as shown in Figure 4. It is a remarkable fact that $\overline{\langle \mathcal{B} \rangle}$ is quite robust against error in the control of atomic times of flight. Besides, the nonlocality of the state generated under these imperfections is revealed even near the weak coupling regime where the dissipation constants are comparable to the coupling constants. To be more specific, $\overline{\langle \mathcal{B} \rangle} \leq 2$ for decay constants only smaller than $\gamma \approx 0.4\lambda$.

## V. CAVITY QED SETUPS

Since some protocols presented in this work are supported by the ability of willingly choosing magnitudes and phases of atom-field coupling constants, we need now to discuss these assumptions on potential physical setups.

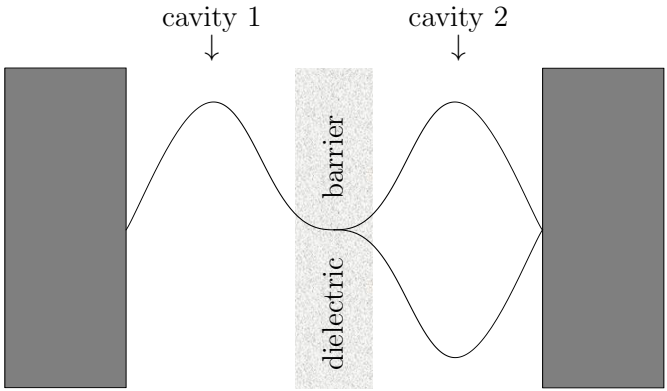

FIG. 5. Sketch of the setup proposed in [8] showing cavities 1 and 2 and the dielectric barrier. The degenerate eigenfunctions are pictorially depicted.

### A. Proposition I

One possible setup consists of two weakly coupled identical cavities 1 and 2 and flying atoms [8]. These cavities are actually the two halves of a perfectly conducting rectangular resonator divided by an insulating barrier of suitable thickness and dielectric constant. The experimenter is then endowed with the choice about which cavity each atom will be sent through. The special feature of this system that allows one to willingly chose the coupling constant phases is that two weakly coupled identical cavities possess two nearly degenerate eigenfunctions (frequency split is approximately proportional to the amplitude transmissivity of the barrier): one symmetric and the other antisymmetric, with respect to the junction of the cavities, as shown in Figure 5. Consequently, when sent to cavity 1, an atom interacts with two positively-valued mode functions, and when sent to cavity 2 it interacts with modes having opposite values. This fits perfectly the needs for phase control found in our state generation protocols.

### B. Proposition II

Now we consider a one-dimensional model describing the interaction of a two-level atom $j$ and $M$ field modes inside a cavity [76]. Such a simplified version of the general two- or three-dimensional model still reveals the essential features of the dynamics [76, 77]. Under the dipole and rotating-wave approximations, the interaction Hamiltonian can be cast in the form

$$H_{\text{int}} = \sum_{n}^{M} g_n^{(j)} (a_n^\dagger \sigma_j^- + a_n \sigma_j^+), \quad (55)$$

where the position-dependent coupling constant for the atom $j$ and mode $n$ is given by

$$g_n^{(j)} = \sqrt{\frac{\omega_n}{\epsilon_0 L}} d_{eg}^{(j)} \sin(k_n r_j), \quad (56)$$

with $k_n = \omega_n/c = n\pi/L$ and $d_{eg}^{(j)}$ denoting the dipole matrix element of atom $j$. Once we are dealing with identical atoms, we may suppose equal values for any atomic dipole matrix elements $d_{eg}^{(j)}$. Under appropriate choices of polarization and by considering the atomic frequency midway the separation of the natural frequencies of consecutive cavity modes, say $n$ and $n+1$, the above Hamiltonian may be reduced to effectively contain two modes

$$H_{\text{int}} = g_n^{(j)} (a_n^\dagger \sigma_j^- + a_n \sigma_j^+) + g_{n+1}^{(j)} (a_{n+1}^\dagger \sigma_j^- + a_{n+1} \sigma_j^+). \quad (57)$$

The problem consists in finding solutions for equations of the kind

$$\sqrt{\tilde{k}_n} \sin[\tilde{k}_n(\tilde{r}_j + 1/2)] = \pm \sqrt{\tilde{k}_{n+1}} \sin[\tilde{k}_{n+1}(\tilde{r}_j + 1/2)], \quad (58)$$

for any atom $j$ and two selected modes, where we have defined parameters scaled by the cavity length, i.e. $\tilde{k}_n = L k_n = n\pi$ and $\tilde{r}_j = r_j/L$ and set the origin $\tilde{r}_j = 0$ to be the center of the cavity, i.e. the cavity mirrors are located at $-L/2$ and $L/2$. From now on, we will consider the lowest order modes 1 and 2 to illustrate the general idea.

Let us consider first the scheme in Section II A, where one atom ($j = 1$) interacts with each of two modes with coupling constants with either equal or opposite phases. In this case, we must solve the simple equation

$$\sqrt{\tilde{k}_1} \sin[\tilde{k}_1(\tilde{r}_1 + 1/2)] = \pm \sqrt{\tilde{k}_2} \sin[\tilde{k}_2(\tilde{r}_1 + 1/2)]. \quad (59)$$

It is easy to find that $\tilde{r}_1 = \mp \frac{1}{\pi} \arcsin(\frac{1}{2\sqrt{2}})$, with the minus referring to the case with equal phases and the plus the other way round. Of course, other positions for the atom may be obtained if we select modes with higher $n$ and $n+1$.

For the schemes with two or more atoms crossing simultaneously the cavity (Sec. refsec:disp), we have to find solutions of system of equations involving terms of the kind (58). In the case of two atoms there is a solution valid for equal couplings with opposite phases [$\tilde{r}_1 = \frac{1}{\pi} \arcsin(\frac{1}{2\sqrt{2}})$ and $\tilde{r}_2 = -\frac{1}{\pi} \arcsin(\frac{1}{2\sqrt{2}})$]. However, in contrast with the first proposition, any other solutions are not physical for this setup . The reason is that they imply putting more than one atom at the same point in the cavity axis. This is a very unfavorable situation because, even if we tolerated some error in the magnitude of the coupling constant by considering both atoms at slightly equal positions, we would still be in trouble because direct dipole-dipole coupling could not be neglected. This would alter the Hamiltonians used in the protocols in a very substantial way. There might still be a solution for this problem in the 2D and 3D versions of the model, but we leave this possibility for future investigations.

## VI. CONCLUSION

We first considered the quasi resonant interaction of two modes with either one qubit at time or with $N$ qubits. In the one excitation sector of the system Hamiltonian, we found the deterministic generation of bosonic Bell states between two modes. W states may also be generated either among the two modes and the qubits or among the $N$ qubits. However, when considering the simultaneous interaction of the modes with the qubits, we observed that the generation of W states occurs only for $N = 2$ or $N = 4$. To circumvent this constraint, we slightly modified the scheme by first preparing the modes in a Bell state. Thus, with the excitation initially stored in this entangled state, $W_{N+2}$ and $W_N$ states are generated for all $N$. Also, we observed that the generation of these states becomes faster with the increase of $N$.

In the far off resonance interaction, we analyze the effective dynamics when two non-degenerate modes interact simultaneously with $N$ qubits. When only one excitation is present, it is possible to generate a wide class of W states involving $N$ qubits, and again the interaction time decreases with $N$. We also observed the generation of GHZ states among three qubits initially prepared in coherent superpositions of spin down and spin up states. With two excitations in the system, we showed the one-step generation of cluster states among four qubits.

We would like to remark that all the protocols for generation of multipartite entangled states presented here are deterministic (no measurements). This became possible due to quantum interferences arising from different choices of coupling constant phases.

We also studied these generation schemes under the the presence of important sources of imperfections such as energy relaxation and fluctuations in the time of flight of the atoms through the cavity. The results showed good fidelities for an appreciable range of the parameters involved. As an application of our generation protocols, we investigated nonlocality tests for the cluster state in the presence of those imperfections. The results are also very robust showing violation of the SASA inequality for dissipative rates very close to the weak coupling regime of cavity QED. The quest for quantum effects in typically unfavorable situations (like the weak coupling regime) is a very active research area [78]. The violation of SASA inequality under such severe imprecisions makes our protocols useful for this kind of investigation.

### ACKNOWLEDGMENTS

P.P.M. thanks the financial support from Coordenação de Aperfeiçoamento de Pessoal de Nível Superior (CAPES) and Conselho Nacional de Desenvolvimento Científico e Tecnológico (CNPq), under Grant N. PNPD0030082 (Programa Nacional de Pós-Doutorado). F.L.S. thanks CNPq for partial support under Grant N. 303042/2008-7. We also acknowledge partial support of the Brazilian National Institutes of Science and Technology of Quantum Information (INCT-IQ).

[1] E. T. Jaynes and F. W. Cummings, Proc. IEEE, **51**, 89 (1963).
[2] M. A. Nielsen and I. L. Chuang, *Quantum Computation and Quantum Information*, 1st ed. (Cambridge University Press, Cambridge, 2000).
[3] B. W. Shore and P. L. Knight, J. Mod. Opt., **40**, 1195 (1993).
[4] A. Messina, S. Maniscalco, and A. Napoli, J. Mod. Opt., **50**, 1 (2003).
[5] G. Benivegna and A. Messina, Phys. Rev. A, **37**, 4747 (1988).
[6] P. Bertet, S. Osnaghi, P. Milman, A. Auffeves, P. Maioli, M. Brune, J.-M. Raimond, and S. Haroche, Phys. Rev. Lett., **88**, 143601 (2002).
[7] S. Gleyzes, S. Kuhr, C. Guerlin, J. Bernu, S. Deléglise, U. B. Hoff, M. Brune, J.-M. Raimond, and S. Haroche, Nature, **446**, 297 (2007).
[8] M. Škarja, N. Mankoč Borštnik, M. Löffler, and H. Walther, Phys. Rev. A, **60**, 3229 (1999).
[9] A. Rauschenbeutel, P. Bertet, S. Osnaghi, G. Nogues, M. Brune, J.-M. Raimond, and S. Haroche, Phys. Rev. A, **64**, 050301 (2001).
[10] D. Gonţa, S. Fritzsche, and T. Radtke, Phys. Rev. A, **77**, 062312 (2008).
[11] D. Gonţa, T. Radtke, and S. Fritzsche, Phys. Rev. A, **79**, 062319 (2009).
[12] D. Gonţa and S. Fritzsche, J. Phys. B, **41**, 095503 (2008).
[13] H. J. Briegel and R. Raussendorf, Phys. Rev. Lett., **86**, 910 (2001).
[14] L. Ye and G.-C. Guo, Phys. Lett. A, **361**, 460 (2007).
[15] L. Ye, Eur. Phys. J. D, **41**, 413 (2007).
[16] A. Blais, J. Gambetta, A. Wallraff, D. I. Schuster, S. M. Girvin, M. H. Devoret, and R. J. Schoelkopf, Phys. Rev. A, **75**, 032329 (2007).
[17] D. Leibfried, R. Blatt, C. Monroe, and D. Wineland, Rev. Mod. Phys., **75**, 281 (2003).
[18] R. Rossi, Jr., A. R. B. de Magalhães, and M. C. Nemes, Physica A, **365**, 402 (2006).
[19] A. R. B. de Magalhães and M. C. Nemes, Phys. Rev. A, **70**, 053825 (2004).
[20] Z. Ficek and S. Swain, *Quantum Interference and Coherence: Theory and Experiments*, 1st ed. (Springer, 2005).
[21] P. A. Ivanov, N. V. Vitanov, and M. B. Plenio, Phys. Rev. A, **78**, 012323 (2008).
[22] M. Ikram and F. Saif, Phys. Rev. A, **66**, 014304 (2002).
[23] Z.-X. Man, F. Su, and Y.-J. Xia, Chin. Sci. Bull., **53**, 2410 (2008).
[24] L. Davidovich, N. Zagury, M. Brune, J.-M. Raimond, and S. Haroche, Phys. Rev. A, **50**, R895 (1994).
[25] P. Agrawal and A. Pati, Phys. Rev. A, **74**, 062320 (2006).
[26] Z. J. Deng, M. Feng, and K. L. Gao, Phys. Rev. A, **73**, 014302 (2006).
[27] J.-G. Li, J. Zou, J.-F. Cai, and B. Shao, Phys. Lett. A, **361**, 59 (2007).
[28] X. Wang, Phys. Rev. A, **64**, 012313 (2001).
[29] S.-B. Zheng, Phys. Rev. A, **70**, 045804 (2004).
[30] X.-J. Zheng, X.-M. Xu, F.-P. Ouyang, and H. Xu, Chin. Phys. Lett., **26**, 040304 (2009).
[31] W. Dür, G. Vidal, and J. I. Cirac, Phys. Rev. A, **62**, 062314 (2000).
[32] G.-P. Guo, C.-F. Li, J. Li, and G.-C. Guo, Phys. Rev. A, **65**, 042102 (2002).
[33] A.-S. Zheng, X.-F. Shen, J.-B. Liu, B. Jie, and Q.-J. Du, Chin. Phys. Lett., **25**, 1195 (2008).
[34] S.-B. Zheng, Chin. Phys., **14**, 533 (2005).
[35] S.-B. Zheng, J. Opt. B, **7**, 10 (2005).
[36] R.-C. Yang, H.-C. Li, X. Lin, Z.-P. Huang, H. Xie, J.-F. Lin, and G.-R. Huang, Opt. Commun., **279**, 399 (2007).
[37] E. M. Becerra-Castro, W. B. Cardoso, A. T. Avelar, and B. Baseia, J. Phys. B, **41**, 215505 (2008).
[38] X.-W. Wang, Int. J. Quantum Inf., **7**, 493 (2009).
[39] F. Dell'Anno, S. De Siena, and F. Illuminati, Phys. Rep., **428**, 53 (2006).

[40] H.-J. Lee, J. Kor. Phys. Soc., **54**, 1369 (2009).
[41] N. Kiesel, M. Bourennane, C. Kurtsiefer, H. Weinfurter, D. Kaszlikowski, W. Laskowski, and M. Zukowski, J. Mod. Opt., **50**, 1131 (2003).
[42] M. Eibl, N. Kiesel, M. Bourennane, C. Kurtsiefer, and H. Weinfurter, Phys. Rev. Lett., **92**, 077901 (2004).
[43] C. F. Roos, M. Riebe, H. Haffner, W. Hansel, J. Benhelm, G. P. T. Lancaster, C. Becher, F. Schmidt-Kaler, and R. Blatt, Science, **304**, 1478 (2004).
[44] B. P. Lanyon and N. K. Langford, New J. Phys., **11**, 013008 (2009).
[45] H. Häffner, W. Hänsel, C. F. Roos, J. Benhelm, D. Chek-al kar, M. Chwalla, T. Körber, U. D. Rapol, M. Riebe, P. O. Schmidt, C. Becher, O. Gühne, W. Dür, and R. Blatt, Nature, **438**, 643 (2005).
[46] S.-B. Zheng and G.-C. Guo, Phys. Rev. Lett., **85**, 2392 (2000).
[47] F. L. Semião, J. Phys. B, **41**, 081004 (2008).
[48] X. B. Zou, K. Pahlke, and W. Mathis, Eur. Phys. J. D, **33**, 297 (2005).
[49] M. Hein, J. Eisert, and H. J. Briegel, Phys. Rev. A, **69**, 062311 (2004).
[50] M. Van den Nest, J. Dehaene, and B. De Moor, Phys. Rev. A, **69**, 022316 (2004).
[51] M. Van den Nest, J. Dehaene, and B. De Moor, Phys. Rev. A, **70**, 034302 (2004).
[52] D. M. Greenberger, M. A. Horne, A. Shimony, and A. Zeilinger, Am. J. Phys., **58**, 1131 (1990).
[53] J. I. Cirac and P. Zoller, Phys. Rev. A, **50**, R2799 (1994).
[54] K. Mølmer and A. Sørensen, Phys. Rev. Lett., **82**, 1835 (1999).
[55] S.-B. Zheng, Opt. Commun., **171**, 77 (1999).
[56] A. Biswas and G. S. Agarwal, J. Mod. Opt., **51**, 1627 (2004).
[57] R. J. Nelson, D. G. Cory, and S. Lloyd, Phys. Rev. A, **61**, 022106 (2000).
[58] D. Bouwmeester, J.-W. Pan, M. Daniell, H. Weinfurter, and A. Zeilinger, Phys. Rev. Lett., **82**, 1345 (1999).
[59] J.-W. Pan, D. Bouwmeester, M. Daniell, H. Weinfurter, and A. Zeilinger, Nature, **403**, 515 (2000).
[60] A. Rauschenbeutel, G. Nogues, S. Osnaghi, P. Bertet, M. Brune, J.-M. Raimond, and S. Haroche, Science, **288**, 2024 (2000).
[61] Y. Hasegawa, G. Badurek, S. Filipp, J. Klepp, R. Loidl, S. Sponar, and H. Rauch, Nucl. Instr. Meth. Phys. Res. Sect. A, **611**, 310 (2009).
[62] R. Raussendorf and H. J. Briegel, Phys. Rev. Lett., **86**, 5188 (2001).
[63] X. B. Zou and W. Mathis, Phys. Rev. A, **72**, 013809 (2005).
[64] X.-H. Zheng, P. Dong, Z.-Y. Xue, and Z.-L. Cao, Phys. Lett. A, **365**, 156 (2007).
[65] R.-C. Yang, H.-C. Li, M.-X. Chen, and X. Lin, Chin. Phys., **15**, 2315 (2006).
[66] X.-Q. Shao and S. Zhang, Chin. Phys. Lett., **25**, 3132 (2008).
[67] Y.-Q. Zhang, S. Zhang, and X.-R. Jin, Phys. Scr., **78**, 065401 (2008).
[68] A.-N. Zhang, C.-Y. Lu, X.-Q. Zhou, Y.-A. Chen, Z. Zhao, T. Yang, and J.-W. Pan, Phys. Rev. A, **73**, 022330 (2006).
[69] C.-Y. Lu, X.-Q. Zhou, O. Guehne, W.-B. Gao, J. Zhang, Z.-S. Yuan, A. Goebel, T. Yang, and J.-W. Pan, Nature Phys., **3**, 91 (2007).
[70] Y. Tokunaga, S. Kuwashiro, T. Yamamoto, M. Koashi, and N. Imoto, Phys. Rev. Lett., **100**, 210501 (2008).
[71] W.-B. Gao, X.-Q. Zhou, J. Zhang, T. Yang, and J.-W. Pan, New J. Phys., **10**, 055003 (2008).
[72] H. J. Carmichael, *An Open Systems Approach to Quantum Optics*, 1st ed. (Springer-Verlag, Berlin Heidelberg, 1993).
[73] V. Scarani, A. Acín, E. Schenck, and M. Aspelmeyer, Phys. Rev. A, **71**, 042325 (2005).
[74] P. Walther, M. Aspelmeyer, K. J. Resch, and A. Zeilinger, Phys. Rev. Lett., **95**, 020403 (2005).
[75] N. Kiesel, C. Schmid, U. Weber, G. Tóth, O. Gühne, R. Ursin, and H. Weinfurter, Phys. Rev. Lett., **95**, 210502 (2005).
[76] V. Bužek, G. Drobný, M. G. Kim, M. Havukainen, and P. L. Knight, Phys. Rev. A, **60**, 582 (1999).
[77] M. Havukainen, G. Drobný, S. Stenholm, and V. Bužek, J. Mod. Opt., **46**, 1343 (1999).
[78] G. Mc Keown, F. L. Semião, H. Jeong, and M. Paternostro, To appear in Phys. Rev. A (2010), e-print arXiv:1004.1316v1 [quant-ph].